\documentclass[twocolumn, aps, prl, notitlepage, nofootinbib]{revtex4-1}
\newcommand{\figmult}{1.}
\usepackage{graphicx}
\usepackage{placeins}
\usepackage{subcaption}
\renewcommand{\vec}[1]{\mathbf{#1}}
\usepackage{amsmath}
\usepackage{braket}
\begin{document}
\title{First-Principles Bulk-Layer Model for Dielectric and Piezoelectric Responses in Superlattices}
\author{J. Bonini, J. W. Bennett, P. Chandra, and K. M. Rabe}

\affiliation{Department of Physics and Astronomy \\ Rutgers University, Piscataway, NJ 08854}
\begin{abstract}
 In the first-principles bulk-layer model the superlattice structure and
polarization are determined by first-principles computation of the bulk
responses of the constituents to the electrical and mechanical boundary
conditions in an insulating superlattice. In this work the model is extended 
to predict functional properties, specifically dielectric permittivity and
piezoelectric response.
A detailed comparison between the bulk-layer model and full first-principles
calculations for three sets of perovskite oxide superlattices,
PbTiO$_3$/BaTiO$_3$, BaTiO$_3$/SrTiO$_3$ and PbTiO$_3$/SrTiO$_3$, is presented.
The bulk-layer model is shown to give an excellent first approximation to these
important functional properties, and to allow for the identification and
investigation of additional physics, including interface reconstruction and
finite size effects.
Technical issues in the generation of the necessary data for constituent
compounds are addressed.
These results form the foundation for a powerful data-driven method to
facilitate discovery and design of
superlattice systems with enhanced and tunable polarization,
dielectric permittivity, and piezoelectric response.
\end{abstract}
\maketitle

\marginparwidth 2.5in
\marginparsep 0.5in
\def\pc#1{\marginpar{\small PC: #1}}
\def\jrb#1{\marginpar{\small JRB: #1}}
\def\kmr#1{\marginpar{\small KMR: #1}}
\def\scr{\scriptsize}


Perovskite oxide superlattices continue to be of both
fundamental and technological interest due to their wide variety of
functional properties as well as the progress in atomic scale precision growth
techniques that
enable their realization
\cite{Schlom_2007, Junquera_2008, dawber_bousquet_2013, Yang2018}.
There is particular interest in systems in which the layering gives rise to
distinctive functional properties, including enhancement of properties such as
the piezoelectric response over those of either constituent \cite{Cooper_2009}.
While the microscopic origins of such behavior could include symmetry breaking
by artificial structuring, a high density of atomically and electronically
reconstructed interfaces, and finite size effects in the unit-cell-scale
constituent layers, early
experimental and first-principles investigation of
$\mathrm{BaTiO}_3$/$\mathrm{SrTiO}_3$ superlattices suggested that the properties of
superlattices, even with ultrashort periods, can in fact be largely predicted by
a ``bulk-layer'' model in which the properties of the superlattice are obtained by
considering the bulk response to the changes in
mechanical and electrical boundary conditions imposed on each constituent layer
by lattice matching and approximate polarization matching \cite{Neaton_2003,
  Tian_2006, Urtiev_2007, PhysRevB.81.144118}.

For a given constituent material, the bulk response to the changes in mechanical
boundary conditions corresponding to lattice matching is readily computed in a
first-principles framework via a strained-bulk calculation in which two
lattice vectors of the bulk material are fixed to match the substrate at the
interface plane, and other structural parameters are relaxed \cite{Pertsev_1998, Di_guez_2005}.
The development of first-principles methods allowing the calculation of
structure
and properties in nonzero uniform electric fields \cite{Souza_2002} and the
subsequent recognition of
the displacement field $\vec{D}$ as the fundamental electrostatic variable
\cite{Stengel_2009}
allow a quantitative determination of how a constituent layer responds to
changes in electrical boundary conditions, including a correct description of
nonlinear behavior at high fields.
The use of these first-principles electrical constitutive relations enables a
fully rigorous implementation of the bulk-layer model.

The bulk-layer model has been successfully applied to a number of perovskite
superlattice systems.
For BaTiO$_3$/SrTiO$_3$, it accounts for the observed polarization of the
SrTiO$_3$ layers \cite{Neaton_2003, Tian_2006} and the evolution of the structure and
polarization with epitaxial strain \cite{Johnston_2005, Jiang_2003, Dawber_2005}.
Extension to the case of
perovskite superlattices with ``charge-mismatched'' constituents (for example,
A$^{3+}$B$^{3+}$O$_3$/A$'^{2+}$B$'^{4+}$O$_3$)
\cite{cazorla14_ab_initiod_charg_mismat_ferroel_super} yielded quantitative
predictions for the epitaxial strain dependence of the structure and
polarization of PbTiO$_3$/BiFeO$_3$ superlattices
\cite{cazorla14_ab_initiod_charg_mismat_ferroel_super, yang12_epitax_short_period_pbtio3_studied}.
For a broader range of superlattice systems, the predictions of the bulk-layer model can be
expected to provide a good starting point from which interface and finite size
effects can be identified and analyzed.

In this Letter, we show how to extend this definitive implementation of the
bulk-layer model to the prediction of dielectric and piezoelectric responses in
insulating superlattices.
For three prototypical titanate superlattice systems, $\mathrm{PbTiO_3/BaTiO_3}$,
$\mathrm{BaTiO_3/SrTiO_3}$, and $\mathrm{PbTiO_3/SrTiO_3}$, we generate the
necessary information about the bulk constituent compounds, apply the bulk-layer
model to the prediction of superlattice structure, polarization, dielectric and
piezoelectric responses and show that the model can accurately predict the
values computed for individual superlattices using full first-principles
methods. Thus, using only a database of computed bulk constituent properties, it
should be possible to map out a large configuration space of superlattice
combinations and investigate the microscopic origins of their functional
properties, leading to a powerful data-driven method to facilitate discovery and
design of superlattice systems with enhanced and tunable polarization,
dielectric permittivity and piezoelectric response.  


The constituent layers of the superlattice are modeled as strained-bulk
materials \cite{Pertsev_1998, Di_guez_2005}
responding uniformly to the changes in mechanical and electrical boundary
conditions produced by the superlattice, specifically lattice matching and
absence of free charge at the interface.  Here,
we consider superlattices epitaxially coherent with a chosen substrate (here, (001)
$\mathrm{SrTiO_3}$), so that the lattice matching is
implemented by fixing two lattice vectors (here, $\vec{a}=(a_0, 0, 0)$ and
$\vec{b}=(0, a_0, 0)$) to match the substrate at the interface plane. The
absence of free charge corresponds to the condition that the displacement field
$\vec{D}$ be uniform throughout the system \cite{Stengel_2009}. Throughout this
Letter we specialize to tetragonal systems where $\vec D$, $\vec E$, and $\vec P$
are along the four-fold axis with magnitudes given by $D$, $E$, and $P$.
For the specified fixed lattice vectors, each constituent material $\alpha$ is
described by the electric-elastic constitutive relations $U(D;\alpha)$,
$c(D;\alpha)$, $E(D;\alpha)$, and $P(D;\alpha)$
corresponding to the energy per unit cell (taken relative to its minimum value),
out-of-plane lattice parameter, electric field, and polarization, respectively.

The systems examined in this Letter are two-component superlattices
with $n_1$ unit cell layers of material $\alpha_1$ and $n_2$ layers of material $\alpha_2$,
with fixed interface charge equal to zero.
The energy of the system is modeled as the sum of the energies of the individual layers:
\begin{equation}
  \label{eqn:U_2}
  U(D) = x U(D;\alpha_1)+ (1-x)U(D;\alpha_2)
\end{equation}
where $x = n_1 / N$ with $N = n_1 + n_2$.
We consider situations in which the voltage drop $V$ across the sample is
controlled, with the $V=0$ short-circuit boundary condition corresponding to the
periodic boundary conditions used in first-principles calculations.
In practice, we first construct
\begin{equation}
  \label{eqn:bound_2}
  \begin{split}
  V(D) &= N x c(D;\alpha_1)E(D;\alpha_1) \\
  &+ N(1-x)c(D;\alpha_2)E(D;\alpha_2)
  \end{split}
\end{equation}

\noindent then the $D$ that corresponds to the target $V$ is obtained by solving
$V(D)=V$ and, if there are multiple solutions, choosing the one that gives
the lowest value of $U(D)$. From this, polarization, out-of-plane lattice
constants, and dielectric and piezoelectric responses can be immediately
obtained as described in the supplemental material. The treatment of more
general superlattices, including more than two components and/or
charge-mismatched constituents, is detailed in the supplemental material. 


\begin{figure}[!h]
\includegraphics[width=\linewidth]{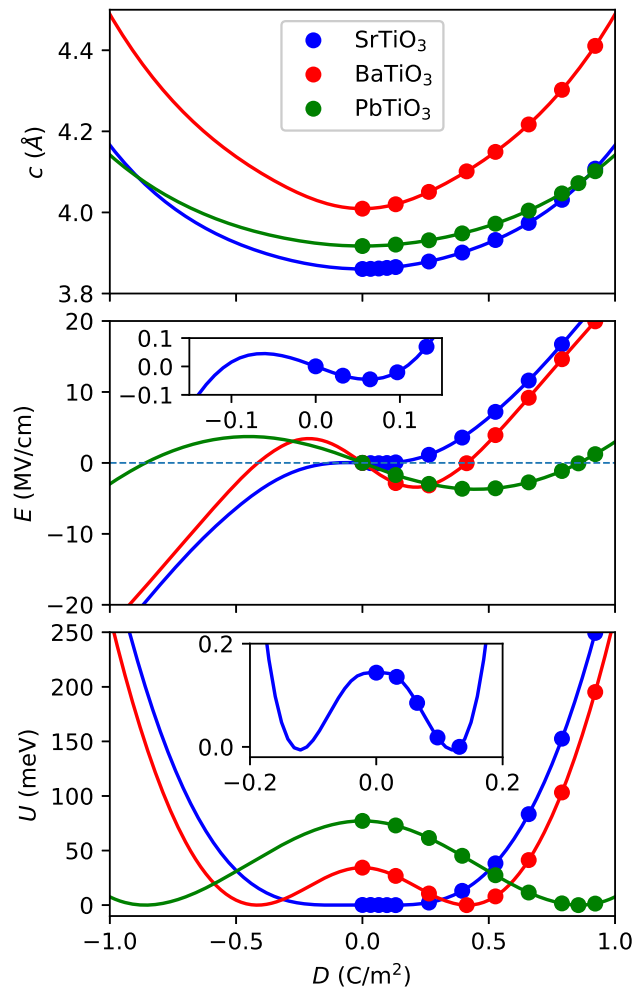}

 \caption{Computed electric-elastic constitutive relations for $\mathrm{SrTiO_3}$,
   $\mathrm{BaTiO_3}$, and $\mathrm{PbTiO_3}$.
   Filled circles show the calculated values and the solid
curves are spline fits. The definite parity of each
function is used to obtain the results for
 negative D.
 The insets zoom in on the slight polar instability computed
 for $\mathrm{SrTiO_3}$.
}
 \label{fig:eos}
\end{figure}

Fig. \ref{fig:eos} shows the electric-elastic constitutive relations for $\mathrm{SrTiO}_3$,
$\mathrm{BaTiO}_3$, and $\mathrm{PbTiO}_3$ computed for displacement fields
ranging from $D=0$
to just above the ground state polarization of $\mathrm{PbTiO}_3$ (
$P=0.85$ $\mathrm{C/m}^2$). The ferroelectrics $\mathrm{BaTiO}_3$ and
$\mathrm{PbTiO}_3$ display a characteristic double
well in the energy and a non-monotonic behavior of the electric field with
displacement field, consistent with the results for $\mathrm{PbTiO}_3$ shown in
\cite{Hong_2011, cazorla15_elect_engin_strain_ferroel_perov}. $\mathrm{SrTiO}_3$
displays its characteristically flat energy 
well and nonlinear evolution of electric field with displacement field
\cite{Antons_2005}, which, as we will discuss below, gives rise to very large
dielectric and piezoelectric responses for superlattices with large $\mathrm{SrTiO}_3$
fraction. Within our first-principles framework, $\mathrm{SrTiO}_3$ is very
slightly polar, with a shallow double well and non-monotonic electric field at
small D as shown in the insets of Fig. \ref{fig:eos}; the experimental observation that
$\mathrm{SrTiO}_3$ is paraelectric down to low temperatures is attributed to the
effects of quantum fluctuations \cite{PhysRevB.19.3593}.
The bulk structural parameters,
polarization, dielectric permittivity, and piezoelectric response for each material are
tabulated in the supplemental material. 


\begin{figure*}[h]
 \begin{subfigure}[]{\textwidth}
   \refstepcounter{subfigure}\label{fig:ptobto}
  \includegraphics[width=\figmult\linewidth]{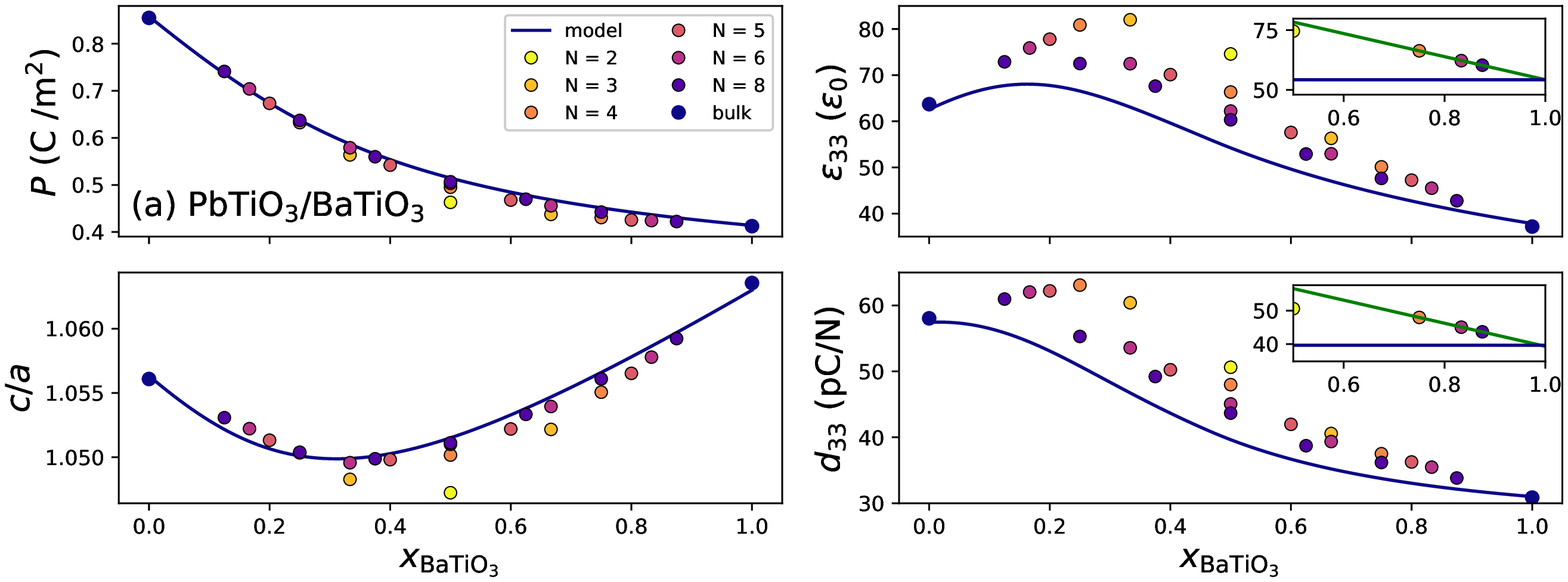}
 \end{subfigure}
\begin{subfigure}[]{\textwidth}
   \refstepcounter{subfigure}\label{fig:btosto}
  \includegraphics[width=\figmult\linewidth]{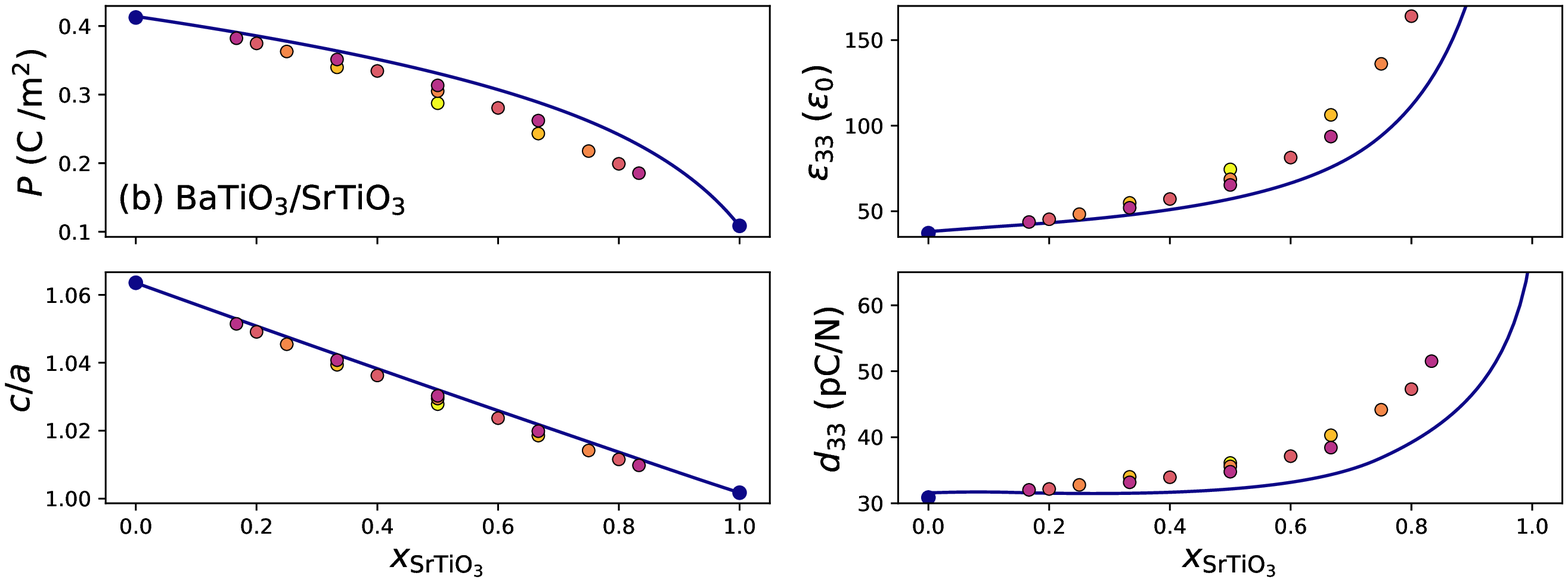}
 \end{subfigure}
 \begin{subfigure}[]{\textwidth}
   \refstepcounter{subfigure}\label{fig:ptosto}
  \includegraphics[width=\figmult\linewidth]{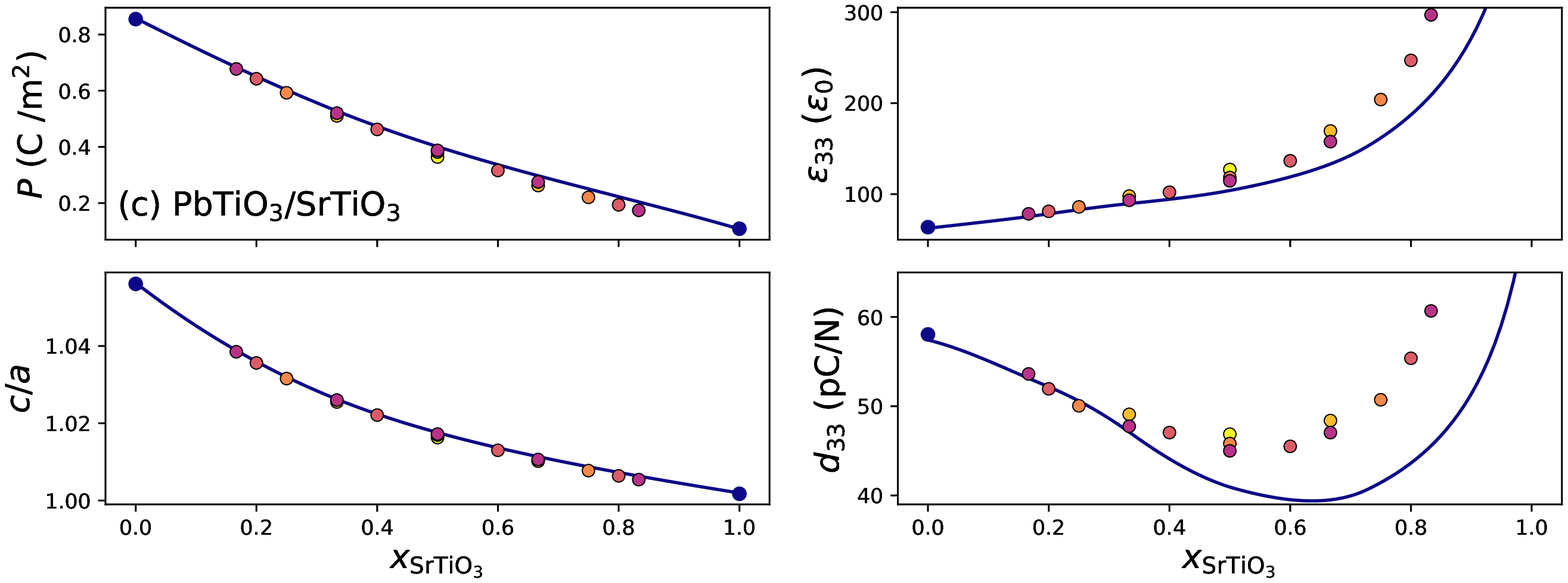}
 \end{subfigure}

 \caption{Spontaneous polarization, tetragonality (c/a), dielectric response
$\epsilon_{33}$ and dielectric response $d_{33}$ for (a) $\mathrm{PbTiO}_3$/$\mathrm{BaTiO}_3$, (b) $\mathrm{BaTiO}_3$/$\mathrm{SrTiO}_3$
and (c) $\mathrm{PbTiO}_3$/$\mathrm{SrTiO}_3$, plotted as functions of the layer fraction $x$ of the
lower polarization constituent.
The bulk-layer model results are shown by a solid line and the first-principles results for
individual superlattices are shown as circles filled by colors corresponding to
the total superlattice period.
The insets in the panels for $\epsilon_{33}$ and $d_{33}$ of $\mathrm{PbTiO}_3$/$\mathrm{BaTiO}_3$ show the
first-principles values for superlattices with $x$ = 0.5 plotted against
$(1 - 1 / N)$, where $N$ is the superlattice period in layers of bulk unit cells,
with a linear fit to the $N>1$ values showing accurate 
convergence to the model value (indicated by the horizontal line).
 The differing scales of the vertical axes in each
 figure are chosen to accommodate the differing ranges over which properties
 vary between systems.
 The imperfect agreement between the end points and the model is discussed in the
 supplemental material.
 }
 \label{fig:model_results}
\end{figure*}

Fig. \ref{fig:model_results} shows the polarization for $\mathrm{PbTiO}_3$/$\mathrm{BaTiO}_3$ superlattices as a
function of $x$, the layer fraction of $\mathrm{BaTiO}_3$. The bulk-layer model shows a bowing
below the linear interpolation between pure $\mathrm{BaTiO}_3$ and pure $\mathrm{PbTiO}_3$.
The first-principles results show only a very weak dependence on the
superlattice period, converging quite rapidly to the model curve with increasing
superlattice period for a given $x$.
The $x$ dependence of model tetragonality $c/a$, where $c=c_{\mathrm{tot}}/N$,
is so strongly bowed that it is nonmonotonic.
Here too, the first-principles results
converge quite rapidly to the model curve with
increasing superlattice period for a given $x$.
The bulk-layer model response functions $\epsilon_{33}$ and $d_{33}$ also show distinctly
nonlinear behavior, with a change in curvature at an intermediate value of
$x$ as well as non-monotonic behavior for $\epsilon_{33}$. The
first-principles results for the response functions show a stronger dependence
on the superlattice period, with substantial enhancement over the model
and with the shortest-period (small $N$), $\mathrm{PbTiO}_3$-richest (small $x$) superlattices displaying
enhancement even above the values of each pure constituent.
With increasing period, these values converge quite accurately to the model, as
illustrated by the insets in Fig. \ref{fig:model_results}.
This is as expected, since the interface and finite size effects in individual
superlattices should become negligible in this limit, and the physics will be
dominated by the effects included in the bulk-layer model, which depends only on
$x$ and is independent of the total superlattice period.

The results for the $\mathrm{BaTiO}_3$/$\mathrm{SrTiO}_3$ superlattices, shown in Fig. \ref{fig:model_results}, show
an upward bowing for the polarization (opposite to that of $\mathrm{PbTiO}_3$/$\mathrm{BaTiO}_3$), and near
linearity for the tetragonality as a function of $x$, the layer fraction of $\mathrm{SrTiO}_3$.
The first principles results show weak dependence on the superlattice period.
The near-flatness of the energy well $U(D; \mathrm{SrTiO}_3)$,
leads to the large dielectric and piezoelectric responses in the $\mathrm{SrTiO}_3$-rich (large
$x$) superlattices. In contrast to $\mathrm{PbTiO}_3$/$\mathrm{BaTiO}_3$ the
first principles results do not converge accurately to the model for large $x$. 

Finally, the results for the $\mathrm{PbTiO}_3$/$\mathrm{SrTiO}_3$ superlattices, shown in Fig.
\ref{fig:model_results}, show only slight bowing for the polarization and the
tetragonality as a function of $x$, the layer fraction of $\mathrm{SrTiO}_3$. The
first-principles results show negligible dependence on superlattice period,
lying on or very close to the model curves even for the shortest-period
superlattices. The dielectric response grows even more rapidly with $x$ than for
$\mathrm{BaTiO}_3$/$\mathrm{SrTiO}_3$ (note the difference in the vertical scale). The piezoelectric response,
in contrast, shows a striking suppression below the pure constituent values at
intermediate values of $x$, which is also clearly evident in the
first-principles results. 


The bowing in the $x$ dependence of the polarization for all three systems can be understood by
considering $x = 0.5$. There, the minimization of $U(D)$ with respect to $D$ requires
$dU(D;\alpha_1)/dD = -dU(D;\alpha_2)/dD$, and examination of Fig. \ref{fig:eos} immediately
shows that the value of $D$, and thus of $P$, that minimizes $U(D)$ is between the
values that minimize the individual $U(D;\alpha_i)$. For the superlattice systems
containing $\mathrm{BaTiO}_3$, the relatively high stiffness of $\mathrm{BaTiO}_3$ around its minimum gives
minimal values of $D$ for $U(D)$ that are closer to that of $\mathrm{BaTiO}_3$ (lower than the
average $D$ for $\mathrm{PbTiO}_3$/$\mathrm{BaTiO}_3$ and higher than the average $D$ for $\mathrm{BaTiO}_3$/$\mathrm{SrTiO}_3$), corresponding
to the observed bowings. The low stiffness of $\mathrm{PbTiO}_3$ combines with the flatness of
$\mathrm{SrTiO}_3$ to give a minimizing $D$ close to and just slightly below the average,
corresponding to the small downward bowing for $\mathrm{PbTiO}_3$/$\mathrm{SrTiO}_3$.

The deviations from the simple linear interpolation values in the tetragonality
($c/a$) can be similarly understood by considering $x=0.5$. In $\mathrm{PbTiO}_3$/$\mathrm{BaTiO}_3$, the value
of $c/a$ computed at the average $D$ of the two constituents ($\bar D$), that is $0.5(c(\bar D ;
\mathrm{PbTiO}_3) + c(\bar D;\mathrm{BaTiO}_3))$ is 4.102 $\AA$, above the linear
interpolation value of 4.087 $\AA$. The downward bowing
in $P$, so that the $D$ at $x=0.5$ is  
well below $\bar D$, is thus completely responsible for lowering the value of
$c/a$ at $x=0.5$ so far as to lead to the nonmonotonic dependence on $x$. In
contrast, for $\mathrm{BaTiO}_3$/$\mathrm{SrTiO}_3$ the upward shift of $c/a$ computed at $\bar D$ relative to
the linear interpolation value is almost equal and opposite in sign to the downward
shift due to the smaller bowing of $P$, so that $c/a$ vs $x$ is almost linear.
Finally, for $\mathrm{PbTiO}_3$/$\mathrm{SrTiO}_3$, the two shifts are comparable in magnitude and both
downward, accounting for the observed downward bowing.

The dependence of the dielectric permittivity and piezoelectric response on
$x$ can similarly be understood as following naturally from the constitutive relations shown in Fig. \ref{fig:eos}.
The details of this analysis, including how the enhancement of $\epsilon_{33}$
in $\mathrm{PbTiO}_3$/$\mathrm{BaTiO}_3$ is related to a supertetragonal phase
of $\mathrm{BaTiO}_3$ and how the suppression of $d_{33}$ in
$\mathrm{PbTiO}_3$/$\mathrm{SrTiO}_3$ results from the negative permittivity
region in $\mathrm{PbTiO}_3$'s constitutive relations, are discussed in the
supplemental material.

An implicit assumption of the bulk-layer model is that the structure within each
constituent layer is uniform.  In the full first-principles calculations, the
structure within each constituent layer is free to vary, and in particular, the
region near the interface can be different from the layer interior. These
additional degrees of freedom,
together with interface effects, contribute to 
the larger responses seen in the full first-principles calculations. This is
particularly pronounced in $\mathrm{BaTiO}_3$/$\mathrm{SrTiO}_3$ and
$\mathrm{PbTiO}_3$/$\mathrm{SrTiO}_3$ superlattices with high $\mathrm{SrTiO}_3$ 
fraction, for which examination of the structure in the $\mathrm{SrTiO}_3$ layer shows
comparatively large variation within the layer, partly accounting for the
discrepancies between the full first-principles superlattice values and the model for
$\epsilon_{33}$ and $d_{33}$.  

In the results presented here, we have considered 5-atom $P4mm$ structures for the
constituent compounds and 1x1x$N$ $P4mm$ structures for the superlattices,
allowing consistent comparisons between the bulk-layer model predictions and the
first-principles calculations. In fact, both experimental and theoretical
investigations of $\mathrm{PbTiO}_3$/$\mathrm{SrTiO}_3$ 
superlattices show that oxygen octahedron rotations appear in the lowest-energy
phases \cite{bousquet08_improp_ferroel_perov_oxide_artif_super, PhysRevB.85.184105, PhysRevB.89.214108}. For comparison to
$\mathrm{PbTiO}_3$/$\mathrm{SrTiO}_3$  experiments, this model therefore should be extended, as done
for $\mathrm{PbTiO}_3/\mathrm{BiFeO}_3$ in \cite{cazorla14_ab_initiod_charg_mismat_ferroel_super}, by laterally
enlarging the unit cells to allow rotations when computing the constitutive
relations. 

In $\mathrm{PbTiO}_3$/$\mathrm{BaTiO}_3$, the dielectric permitivitty and
piezoelectric responses show strong period-dependent enhancements relative to
the bulk-layer model, with the largest enhancements for the shortest period
superlattices: 38\% in $\epsilon_{33}$ for the 1:1 superlattice and 32\% in
$d_{33}$ for the 2:1 superlattice. For both $\epsilon_{33}$ and  $d_{33}$, the
highest values at intermediate $x$ are above the values for either constituent.
This signals the contribution of the interfaces, including atomic and electronic
reconstruction, and finite size effects. Detailed examination of the computed
superlattice structures and phonons could give more information about these
contributions; this is the subject of future work.  

In summary, we have extended the first-principles bulk-layer model, which
predicts the properties of superlattices from the bulk constituent responses
to changing mechanical and electrical boundary conditions, to the prediction of
dielectric and piezoelectric responses in insulating superlattices. We have
presented a quantitative comparison between the model and full first-principles
calculations for three sets of superlattices ($\mathrm{PbTiO}_3$/$\mathrm{BaTiO}_3$, $\mathrm{BaTiO}_3$/$\mathrm{SrTiO}_3$ and $\mathrm{PbTiO}_3$/$\mathrm{SrTiO}_3$)
demonstrating that the model provides an excellent first approximation to the
polarization, tetragonality, dielectric permittivity and piezoelectric response
of these systems allowing the identification of interface and finite-size
effect contributions. Expansion of the constituent database will allow the
efficient exploration of a large configuration space of superlattices, enabling
the data-driven design and discovery of superlattice materials with targeted
functional properties. 

This work is supported by NSF DMR-1334428 and Office of Naval Research 
N00014-17-1-2770. Part of this work was performed at the Aspen Center for
Physics, which is supported by NSF PHY-1607611.
We thank Valentino Cooper, Cyrus Dreyer, Don Hamann, Janice Musfeldt, David
Vanderbilt, and Tahir Yusufaly for useful discussions. We also thank
Ron Cohen for suggesting the modifications to the fixed displacement field 
implementation discussed in the supplemental material. Computing resources were
provided by the ERDC DoD Supercomputing Resource Center. 

\FloatBarrier

\pagebreak
\clearpage
\widetext

\setcounter{equation}{0}
\setcounter{figure}{0}
\setcounter{table}{0}
\setcounter{page}{1}
\renewcommand{\thetable}{S\arabic{table}}
\renewcommand{\theequation}{S\arabic{equation}}
\renewcommand{\thefigure}{S\arabic{figure}}


\begin{center}
\textbf{\large First-Principles Bulk-Layer Model for Dielectric and Piezoelectric 
               Responses in Superlattices: Supplemental Material}
\end{center}

\section{General Formulation of the Model }
For the superlattice consisting of periodic repeats of $k$ layers of unit cell
thickness $n_i$; $i=1,...k$, with superlattice period $N=\sum_i n_i$, the total
energy is taken as the sum of the energies of the individual layers:  
\begin{equation}
  \label{eqn:U_gen}
  U(D) = \sum_i x_i U(D-\sigma_i;\alpha_i)
\end{equation}
where $x_i = n_i / N$ and 
the case of charge-mismatched constituents is treated by including fixed
interface charges $\sigma$ as in
\cite{cazorla14_ab_initiod_charg_mismat_ferroel_super},  so that
$\sigma_i=\sum_{j=1}^{i-1}\sigma_{j,j+1}$, $\sigma_{j,j+1}$ is the fixed
interface charge at the interface between layer $j$ and layer $j+1$, and
$\sigma_1=0$. 

We consider situations in which the voltage drop $V$ across the sample is
controlled, with the $V=0$ short-circuit boundary condition corresponding to the
periodic boundary conditions used in first-principles calculations.
In practice, we first construct
\begin{equation}
  \label{eqn:bound_gen}
  V(D) = \sum_in_iE(D-\sigma_i;\alpha_i)c(D-\sigma_i;\alpha_i)
\end{equation}
\noindent The $D$ that corresponds to the target $V$ is obtained by solving $V(
D)=V$ and if there are multiple solutions, then choosing the one that gives the
lowest value of $U(D)$. For $V=0$, this is equivalent to minimizing
$U(D)$ with respect to $D$ as in
\cite{cazorla14_ab_initiod_charg_mismat_ferroel_super}. We then construct
$c_{\mathrm{tot}}(D)=\sum_in_ic(D - \sigma_i; \alpha_i)$ and
$E_{\mathrm{ext}}(D)=V( D)/c_{\mathrm{tot}}(D)$
and their derivatives with
respect to $D$, from which we obtain the zero-stress dielectric
permittivity $\epsilon_{33} = dD / dE_{\mathrm{ext}} =  (dE_{\mathrm{ext}} / dD)^{-1}$
and the piezoelectric response $d_{33}=c_{\mathrm{tot}}^{-1} (d c_{\mathrm{tot}}
/dD) (dD /d E_{\mathrm{ext}}) = g_{33} \epsilon_{33} $ where $g_{33} =
c_{\mathrm{tot}}^{-1} d c_{\mathrm{tot}}/ d D$. Note that the dielectric and piezoelectric
constants used in this work are for fixed in-plane lattice constants; this is
discussed below in the section ``First-Principles Linear-Response Calculations With Epitaxial Constraints''.
While in the main text we discuss two-component superlattices, the model
as formulated here can be applied to an arbitrary number of components. Results
for three-component $\mathrm{PbTiO}_3$/$\mathrm{SrTiO}_3$/$\mathrm{BaTiO}_3$ systems are shown below.

  \includegraphics[width=0.5\linewidth]{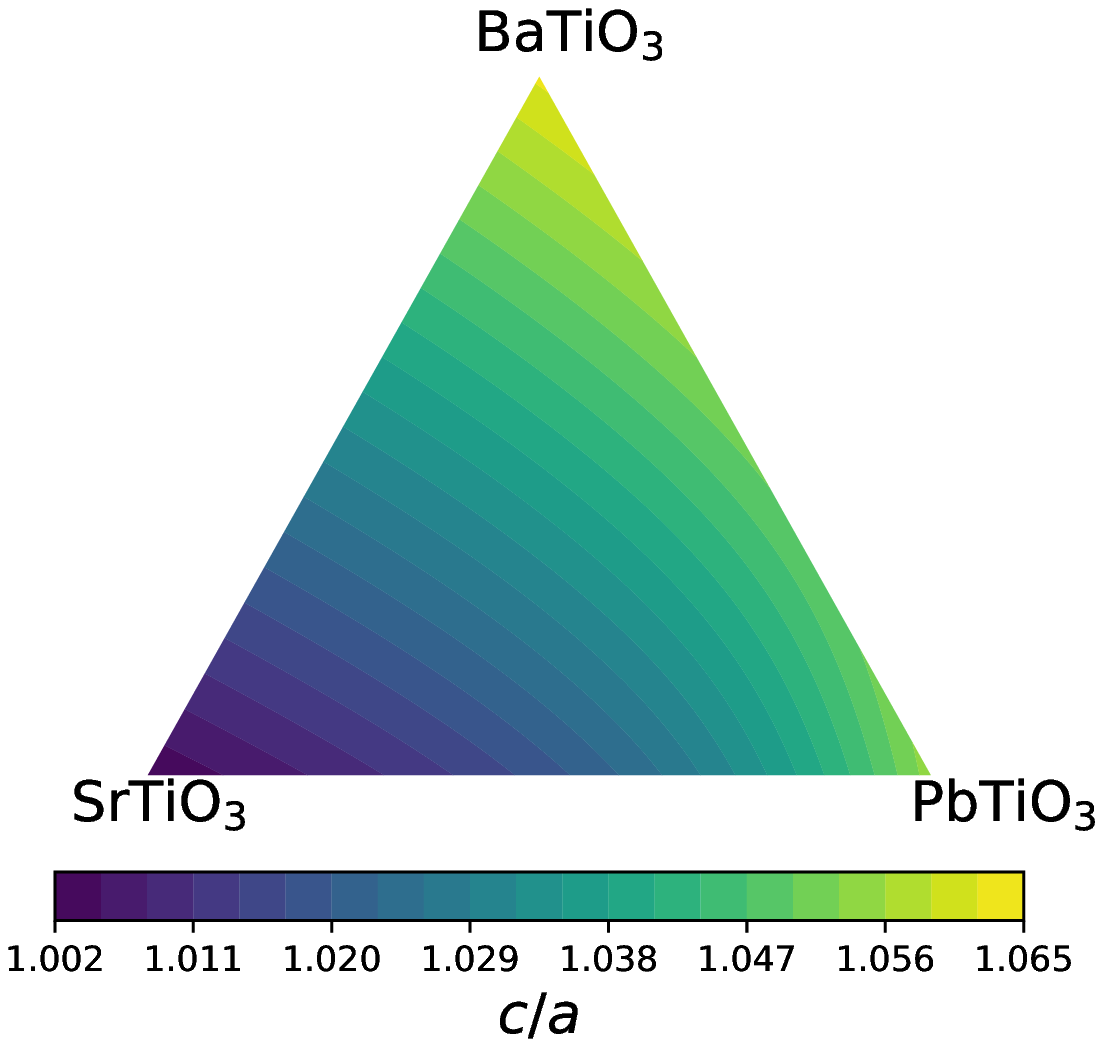}
  \includegraphics[width=0.5\linewidth]{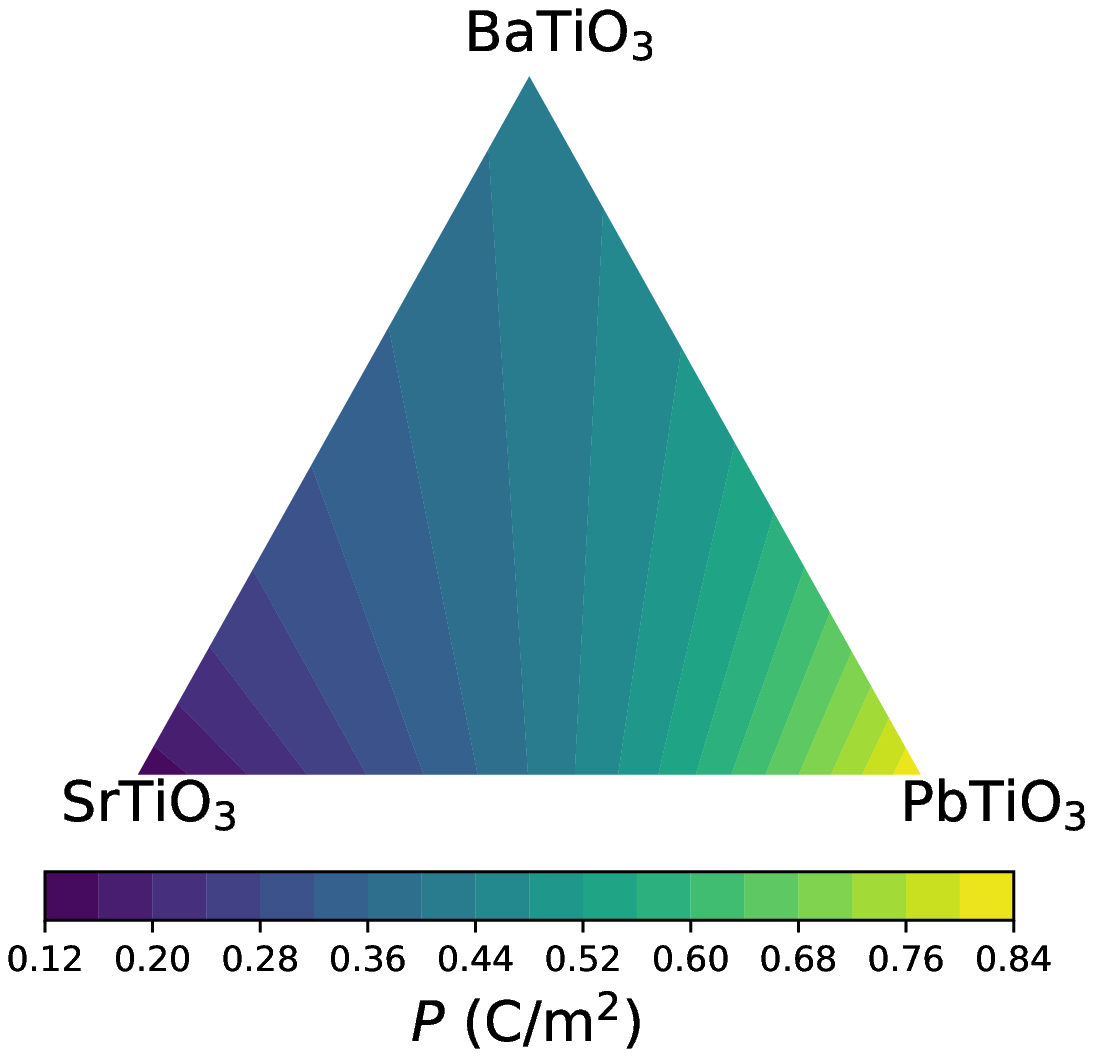}

  \includegraphics[width=0.5\linewidth]{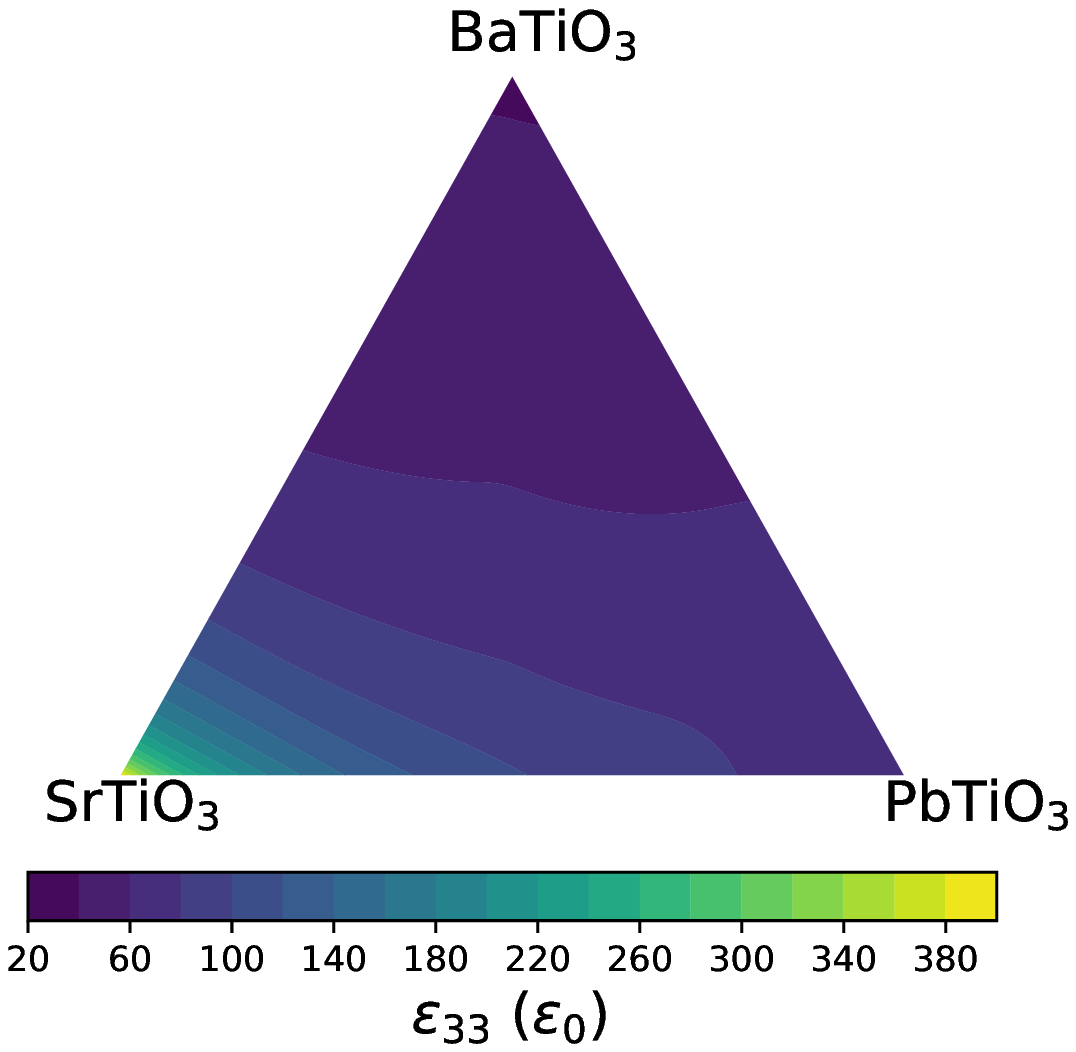}
  \includegraphics[width=0.5\linewidth]{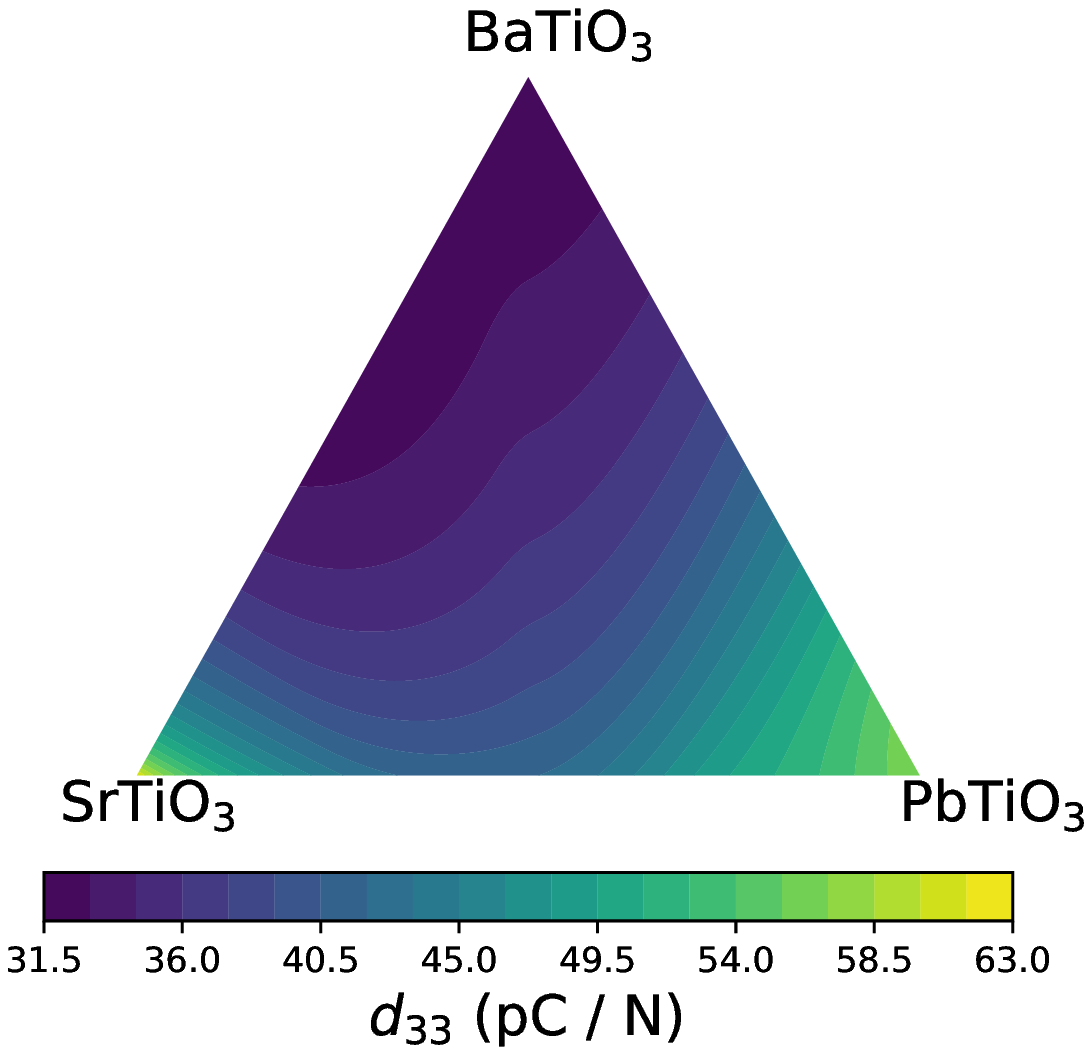}

  \includegraphics[width=0.5\linewidth]{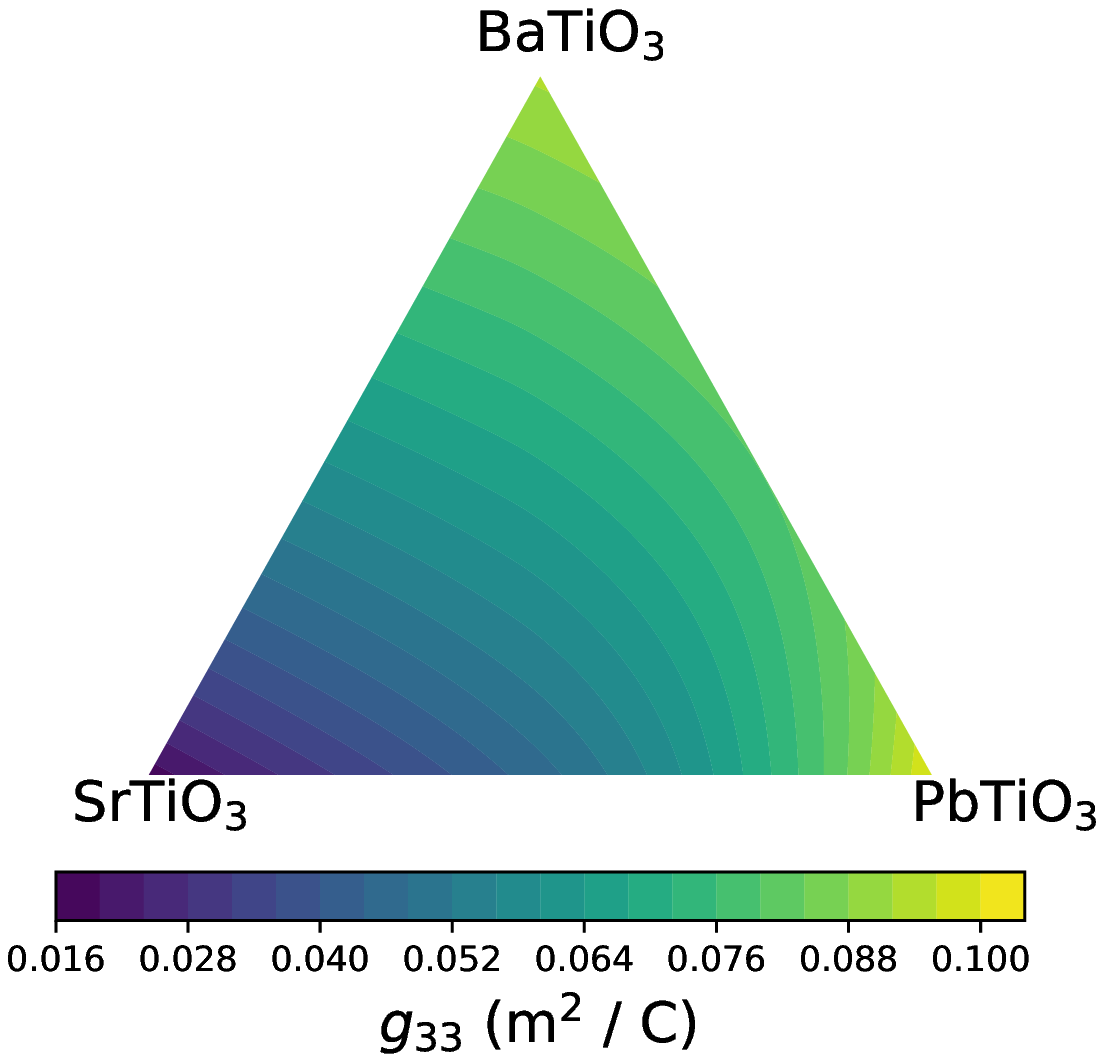}

\section{Determination of electric-elastic constitutive relations}

The nonlinear responses of the constituent layers of the superlattice to changes
in mechanical and electrical boundary conditions are modeled
by electric-elastic constitutive relations $U(\vec{D}; \alpha)$, $\vec
c(\vec{D}; \alpha)$, $\vec{E}(\vec{D}; \alpha)$, and $\vec{P}(\vec{D}; \alpha)$,
where $U$ is the energy, $\vec{c}$ is the out of plane lattice vector, $\vec{E}$
is the electric field, $\vec{P}$ is the polarization, and $\alpha$ denotes the
constituent material.
In this work we consider systems
with symmetry such that $\vec D=(0,0,D)$, $\vec c=(0,0,c)$, $\vec E=(0,0,E)$, and $\vec P=(0,0,P)$
so that the functions reduce to $U(D; \alpha)$, $c(D; \alpha)$
$E(D; \alpha)$, and $P(D; \alpha)$.
To determine these functions in the relevant range of $D$, we perform
first-principles fixed-$D$ calculations as implemented in ABINIT \cite{Gonze20092582, gonze2005brief, Gonze2002478}.
In this approach the energy is given by:
\begin{equation}
  U(D; \alpha) = \min_{\{\vec{r}_i\}} \left[  E_{\mathrm{KS}}(\{\vec{r}_i\}; \alpha) + \frac{\Omega \epsilon_0}{2}(D - P(\{\vec{r}_i)\}; \alpha))^2  \right] 
  \label{eqn:fixed_D}
\end{equation}
where $E_{\mathrm{KS}}$ is the Kohn-Sham
energy functional, $\Omega$ is the unit cell volume,
$\epsilon_0$ is the permittivity of free space, and $P$ is the Berry phase
polarization \cite{spaldin12_begin_guide_to_moder_theor_polar, king-smith93_theor_polar_cryst_solid}.

We have found that for structural relaxation at $D$ much different than the
spontaneous polarization additional care must generally be taken to successfully
converge the calculation. 
In the fixed displacement field implementation in ABINIT the functional (\ref{eqn:fixed_D}) is not minimized directly. Instead
the existing routines for performing fixed electric field ($E$) are utilized (see \cite{Stengel_2009}).
During a single step of structural relaxation the ionic structure is fixed while
the electronic Kohn-Sham wavefunctions are determined by applying varied $E$
fields so that $E = \frac{1}{\epsilon_0}(D-P)$ is satisfied upon convergence.
If the unrelaxed
structure is far from the relaxed structure corresponding to the target $D$, the
ABINIT implementation will fail as the relevant
 $E$ values become so large that the energy 
functional no longer has a minimum as discussed in \cite{Souza_2002}.

One way to avoid this is by choosing starting structures close
to the target structure for a particular $D$ by changing $D$ in
small increments and using the structure from the previous step.
However, we have found that with a small modification \footnote{suggested by R.
  E. Cohen} to the fixed displacement field routines we can avoid this fine-scale
incrementing of $D$, allowing for roughly an order of magnitude increase in
efficiency. The modification is to cap the 
electric field allowed during intermediate ionic steps. This allows the
structure to continue to relax towards structures for which ${P}$ is
closer to the target ${D}$ and the electric field is smaller. At the largest values of
$D$, it might be that the true electric field is larger than the capping value,
yielding results in which the
electric field in the final structure is equal to the capping value. In this
situation either the cap has to be gradually increased (if there is still a minimum of the $U$
function in this range of $E$) or no result can be obtained for $D$ at and above this value.
We have found a capping E field of $5 \times
10^{-3}$ a.u. ($2.57 \times 10^{9}$ V/m) to work well for the materials studied here.
Note that even in an implementation where (\ref{eqn:fixed_D}) was minimized directly a similar
issue would still occur in that there would be no minimum in the energy
functional for large $D-P$, and a similar limit on the second term in equation
(\ref{eqn:fixed_D}) would need to be imposed for intermediate relaxation steps.

While this capping of the electric field allows for relaxation at $ D$ with
starting structures which have a relatively large $ ( D -  P) $,
another issue can arise if this difference is too large.
Since ${P}$ of a periodic system takes values on a lattice, special care
must be taken to choose the correct branch. Since the default behavior is to
choose this branch so as to minimize the internal energy, if one starts a
calculation fixing ${D}$ to a value that differs by a polarization quantum from the
 spontaneous polarization of the starting structure, the ${P}$ will
stay on the wrong branch. This can be avoided by ramping $D$ from its zero field
value using steps smaller than a polarization quantum.
For the systems examined here this step size
is over an order
of magnitude larger than previously required for the calculations performed
with uncapped $E$.

To compute derivatives of the functions $U$, $E$, $P$ and $c$, we use a spline fit to
the first-principles calculations. The relation $E(D) = \frac{1}{\Omega(D)}\frac{dU}{dD}$ is satisified to high
accuracy.

\section{First-Principles Calculation Details}
\begin{table}[!htb]
 \begin{minipage}{0.3\linewidth}
 \begin{tabular}{| l c | c c l |}
 \hline
 \multicolumn{5}{ |c| }{$\mathrm{SrTiO}_3$ $P4mm$ (99)} \\
 \multicolumn{5}{ |c| }{$a=3.857\AA$, $c= 3.864\AA$} \\
 \multicolumn{5}{ |c| }{$P$ = 0.109 C/m$^2$} \\
 \hline
 Sr & 1a & 0 & 0 & 0 \\
 \hline
 Ti & 1b & 1/2 & 1/2 & 0.501 \\
 \hline
 O & 1b & 1/2 & 1/2 & 0.991 \\
 & 2c & 1/2 & 0 & 0.490 \\
 \hline
 \end{tabular}
 \end{minipage}
 \begin{minipage}{0.3\linewidth}
 \begin{tabular}{| l c | c c l |}
 \hline
 \multicolumn{5}{ |c| }{$\mathrm{BaTiO}_3$ $P4mm$ (99)} \\
 \multicolumn{5}{ |c| }{$a=3.857\AA$, $c=4.102\AA$ } \\
 \multicolumn{5}{ |c| }{$P$ = 0.412 C/m$^2$} \\
 \hline
 Ba & 1a & 0 & 0 & 0 \\
 \hline
 Ti & 1b & 1/2 & 1/2 & 0.517 \\
 \hline
 O & 1b & 1/2 & 1/2 & 0.963 \\
 & 2c & 1/2 & 0 & 0.475 \\
 \hline
 \end{tabular}
 \end{minipage}
 \begin{minipage}{0.3\linewidth}
 \begin{tabular}{| l c | c c l |}
 \hline
 \multicolumn{5}{ |c| }{$\mathrm{PbTiO}_3$ $P4mm$ (99)} \\
 \multicolumn{5}{ |c| }{$a=3.857 \AA$, $c=4.073 \AA$ } \\
 \multicolumn{5}{ |c| }{$P$ = 0.855 C/m$^2$} \\
 \hline
 Pb & 1a & 0 & 0 & 0 \\
 \hline
 Ti & 1b & 1/2 & 1/2 & 0.466 \\
 \hline
 O & 1b & 1/2 & 1/2 & 0.903 \\
 & 2c & 1/2 & 0 & 0.392 \\
 \hline
 \end{tabular}
 \end{minipage}
 \caption{Computed structural parameters and polarization ($P$) of each epitaxially
   constrained constituent material.}
 \label{table:structures}
\end{table}
We performed first-principles density-functional-theory calculations with the
local density approximation (LDA) using the ABINIT package \cite{Gonze20092582,
  gonze2005brief, Gonze2002478}. Norm-conserving pseudopotentials were generated 
with the Opium code \cite{Bennett_2012, opium}. An energy cutoff of 800 eV was used with a
10$\times$10$\times$10 Monkhorst-Pack grid to sample the Brillouin zone for
5-atom-unit-cell systems, and equivalent k point densities for the superlattice
systems\cite{PhysRevB.13.5188}. Structural relaxations were performed with a
force threshold of 10 meV/\AA,  except for $\mathrm{SrTiO}_3$ fixed-displacement-field
calculations where the slightly polar structure required a stricter 
convergence of 1 meV/\AA. 
For the superlattices, polarization was computed
using the Berry phase formalism \cite{king-smith93_theor_polar_cryst_solid}, and
dielectric and piezoelectric responses were computed using density functional
perturbation theory (DFPT) \cite{PhysRevB.55.10337, PhysRevB.55.10355, Wu_2005}.
The electric-elastic constitutive relations for the materials $\mathrm{BaTiO}_3$,
$\mathrm{PbTiO}_3$ and $\mathrm{SrTiO}_3$ were computed using fixed displacement field calculations for
the five atom unit cell \cite{Stengel_2009}. Convergence issues encountered (and 
the measures taken to remedy them) in performing the fixed displacement-field
calculations were discussed in the previous section.

\section{Analysis of dielectric permittivity and piezoelectric response of superlattices}
\begin{figure}[h]
  \includegraphics[width=0.5\linewidth]{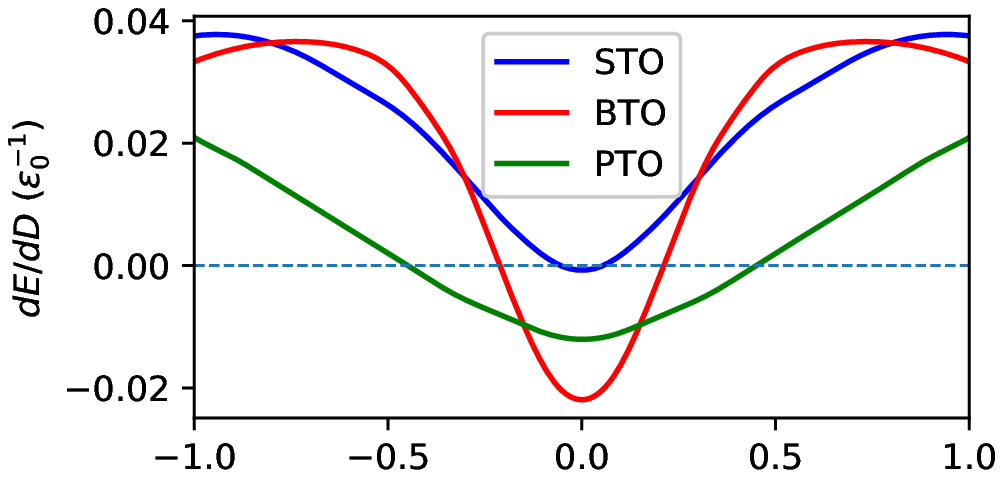}
  \caption{The derivative of the $E(D; \alpha)$ curves (from Fig. 1 of the
    main text) with respect to $D$ for $\mathrm{SrTiO_3}$, $\mathrm{BaTiO_3}$, and $\mathrm{PbTiO_3}$.}
  \label{fig:dEdD}
\end{figure}

\begin{figure}[h]
 \begin{subfigure}[]{\textwidth}
 \caption{$\mathrm{PbTiO}_3$/$\mathrm{BaTiO}_3$}
  \label{fig:ptobtog33}
\includegraphics[width=0.5\linewidth]{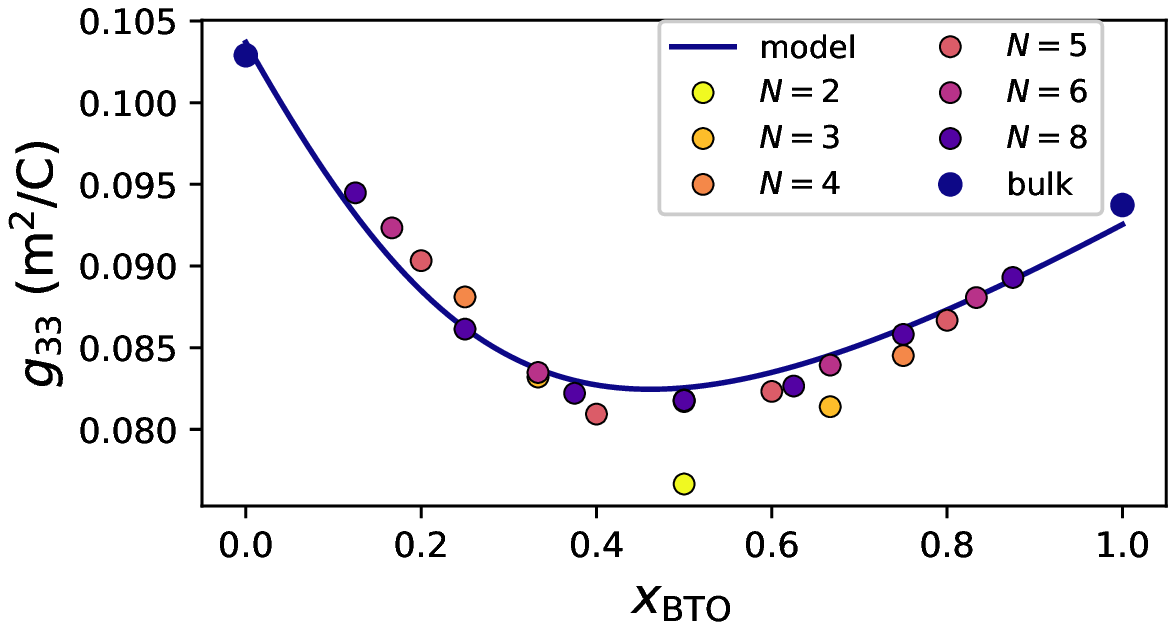}
 \end{subfigure}

 \begin{subfigure}[]{\textwidth}
 \caption{$\mathrm{BaTiO}_3$/$\mathrm{SrTiO}_3$}
  \label{fig:btostog33}
 \includegraphics[width=0.5\linewidth]{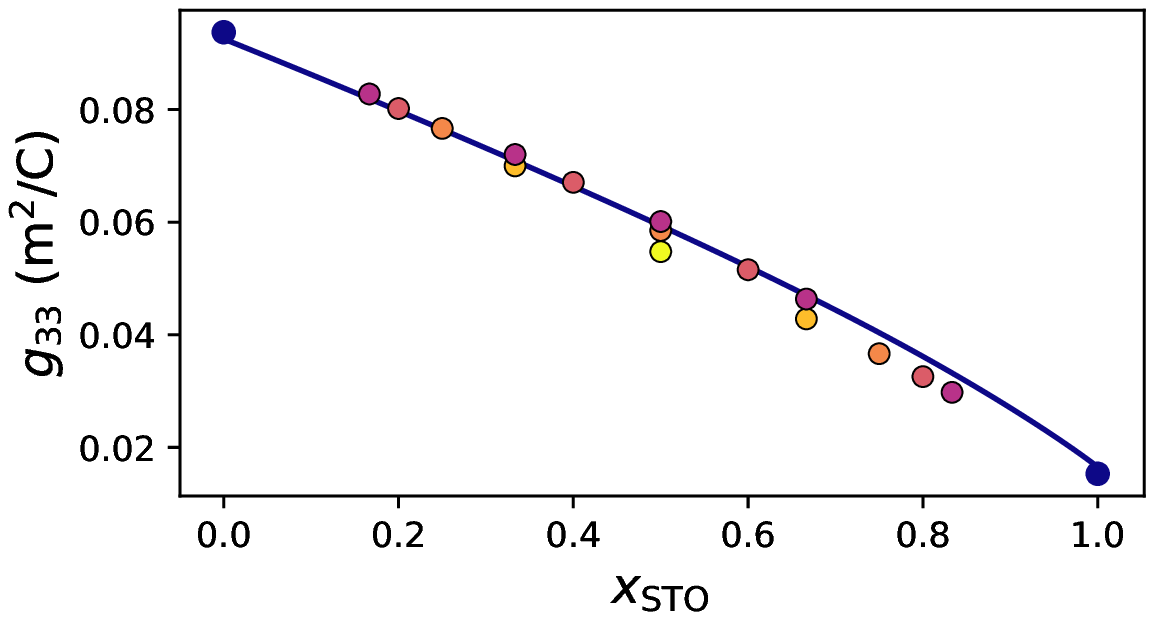}
 \end{subfigure}

 \begin{subfigure}[]{\textwidth}
 \caption{$\mathrm{PbTiO}_3$/$\mathrm{SrTiO}_3$}
  \label{fig:ptostog33}
\includegraphics[width=0.5\linewidth]{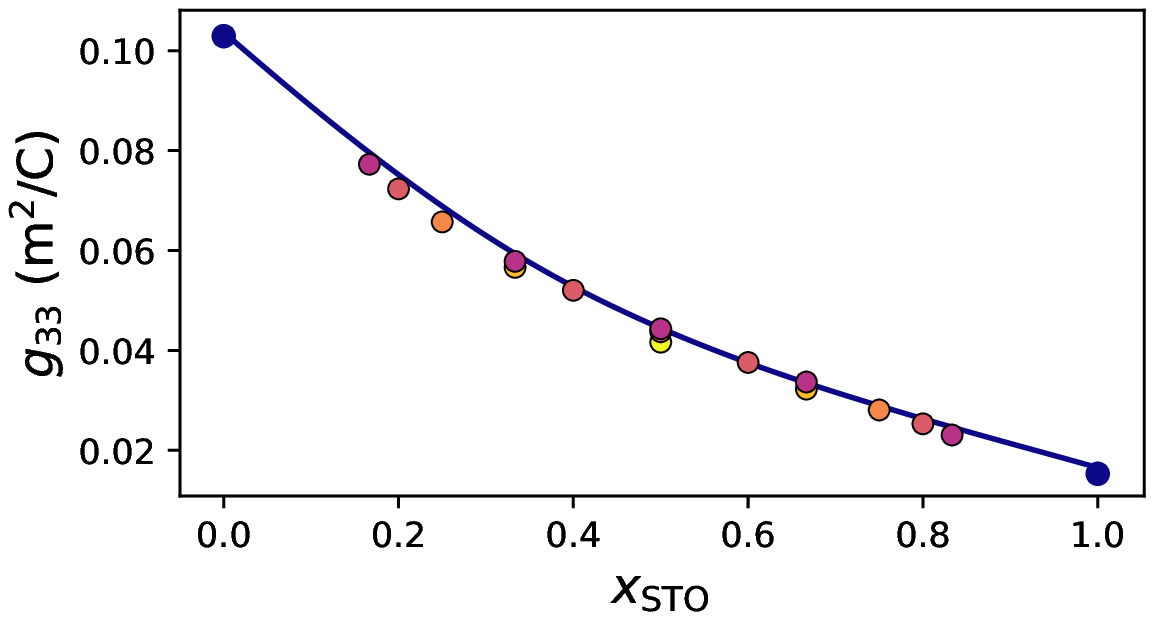}
 \end{subfigure}

 \caption{Model and first principles results for $g_{33}=\frac{1}{c}\frac{dc}{dD}$ for (a)
   $\mathrm{PbTiO}_3$/$\mathrm{BaTiO}_3$, (b) $\mathrm{BaTiO}_3$/$\mathrm{SrTiO}_3$, and (c) $\mathrm{PbTiO}_3$/$\mathrm{SrTiO}_3$ as functions of layer fraction $x$ of
   the lower polarization constituent.
   The bulk-layer model results are shown by a solid line and the first-principles results for
   individual superlattices are shown as circles filled by colors corresponding to
   the total superlattice period.
 }
 \label{fig:g33}
\end{figure}

The dielectric permittivity of the superlattice
$\epsilon_{33}=dD/dE_{\mathrm{ext}}$ can be expressed in terms of the behavior
of individual layers as:
\begin{equation}
  \epsilon_{33} = \frac{\sum_ix_ic(D; \alpha_i)}{\sum_ix_ic(D; \alpha_i)\frac{dE(D; \alpha_i)}{dD}}
  \label{eqn:eps33}
\end{equation}
The non-monotonic behavior of 
$\epsilon_{33}$ in $\mathrm{PbTiO}_3$/$\mathrm{BaTiO}_3$ can be partly
attributed to an anomaly in the high-D behavior of $\mathrm{BaTiO}_3$, with a nonlinear
softening for $D > 0.6 \ \mathrm{C}/\mathrm{m}^2$, evident in 
Fig. \ref{fig:dEdD}.
This softening arises from proximity in the energy
landscape to a highly polar supertetragonal phase of $\mathrm{BaTiO}_3$ which
has been predicted to be stable at large negative pressure \cite{Di_guez_2004,
  tinte03_anomal_enhan_tetrag_inpbt_by_negat_press}.
While the supertetragonal phase is not even metastable under the mechanical and
electrical boundary conditions explored, the values of $D$ achieved in the $\mathrm{BaTiO}_3$
layer in superlattices with a large fraction of $\mathrm{PbTiO}_3$ are in this anomalous
regime. Similarly large values of $D$ are achieved in $\mathrm{SrTiO}_3$ layers for $\mathrm{PbTiO}_3$/$\mathrm{SrTiO}_3$
superlattices with low $\mathrm{SrTiO}_3$ fraction. However, as can be seen in the $\mathrm{SrTiO}_3$ $dE/dD$
curve in Fig. \ref{fig:dEdD}, while $dE/dD$ does begin to soften in $\mathrm{SrTiO}_3$ it
never decreases in the relevant range of $D$. Furthermore, the large
permittivity of $\mathrm{SrTiO}_3$
dominates the evolution of $\epsilon_{33}$ with $x$,
and any enhancement due to effects on
the energy landscape from a supertetragonal phase are comparatively negligible. 
The dielectric permittivity of $\mathrm{PbTiO}_3$/$\mathrm{SrTiO}_3$ is seen to increase more rapidly with $x$ than
that of $\mathrm{BaTiO}_3$/$\mathrm{SrTiO}_3$ (notice the difference in scales between the two plots). While there
is a contribution from the slight softening of $\mathrm{SrTiO}_3$ at high $D$, $\mathrm{PbTiO}_3$/$\mathrm{SrTiO}_3$ is also
the only one of the three systems examined here where 
one of the constituents has a negative $dE(D; \alpha)/dD$ for a large range of
$x$ (see $\mathrm{PbTiO}_3$ in Fig. 1 at $D < 0.55$).
A negative  $dE(D; \alpha_i)/dD$ in the denominator of equation (\ref{eqn:eps33}) increases the
permittivity of the superlattice \cite{PhysRevLett.99.227601}.


The behavior of $d_{33}$ for each system
 can be understood by first
recalling that $d_{33}$ = $\epsilon_{33} g_{33}$. As can be seen in Fig.
\ref{fig:g33}, each system's $g_{33}(x)$ has a bowing following that 
of the polarization bowing for reasons analogous to those discussed regarding
the tetragonality.
In $\mathrm{PbTiO}_3$/$\mathrm{BaTiO}_3$ the downward bowing of $g_{33}(x)$ is so
strong that it is nonmonotonic. When multiplied by $\epsilon_{33}(x)$, which 
has the previously discussed enhancement, the resulting
$d_{33}(x)$ is monotonically decreasing, with a change in curvature.
For both $\mathrm{BaTiO}_3$/$\mathrm{SrTiO}_3$ and $\mathrm{PbTiO}_3$/$\mathrm{SrTiO}_3$ $g_{33}$ is a monotonically
decreasing function of $x$, while $\epsilon_{33}$ is monotonically increasing,
but their $d_{33}$ curves exhibit qualitatively different behavior.
This can be understood by considering how the slope at any given $x$
relates to the slopes and magnitudes of $\epsilon_{33}$ and
$g_{33}$. $$\frac{d(d_{33})}{dx} = \frac{d\epsilon_{33}}{dx}g_{33}(x) +
\epsilon_{33}(x)\frac{dg_{33}}{dx}$$ 
For both $\mathrm{BaTiO}_3$/$\mathrm{SrTiO}_3$ and $\mathrm{PbTiO}_3$/$\mathrm{SrTiO}_3$ the first term is always positive and the
second term is always negative. Then $d_{33}$ will have a negative slope in
regions where the following is satisfied:
$$\frac{1}{g_{33}}|\frac{dg_{33}}{dx}|\epsilon_{33} > \frac{d\epsilon_{33}}{dx}$$ 
For both $\mathrm{BaTiO}_3$/$\mathrm{SrTiO}_3$ and $\mathrm{PbTiO}_3$/$\mathrm{SrTiO}_3$ systems $d\epsilon_{33} / d x_{\mathrm{SrTiO}_3}$ comes to dominate in the large
$x_{\mathrm{SrTiO}_3}$ limit, resulting in the above condition not being satisfied
implying a positive slope at large $x$.
For $\mathrm{BaTiO}_3$/$\mathrm{SrTiO}_3$ the above condition is not satisfied at $x=0$, so $d_{33}(x)$ can
monotonically increase.
In $\mathrm{PbTiO}_3$/$\mathrm{SrTiO}_3$ the larger $\epsilon_{33}$ of $\mathrm{PbTiO}_3$ (discussed above), combined with the positive curvature of
$g_{33}$  result in the above inequality
being satisfied for $x=0$, leading to
the nonmonotonic behavior observed in $d_{33}$ in Fig. 2 of the main text.

\section{First-principles linear-response calculations with epitaxial
constraints}
The dielectric and piezoelectric responses obtained in the model correspond
to the response of the system with in-plane lattice constants fixed to those of
$\mathrm{SrTiO_3}$(001), rather than the zero-stress responses designated
$\epsilon_{33}$ and $d_{33}$ in ABINIT. In this section, we give details on
obtaining the reported responses from the quantities provided by ABINIT.

The epitaxially-constrained dielectric permittivity is
$(\frac{dD_3}{dE_3})_{\sigma_3=0}$ where $\sigma$ is the stress in Voigt
notation.
To obtain $(\frac{dD_3}{dE_3})_{\sigma_3=0}$ from the quantities
provided by ABINIT,
we note that with the condition that the in-plane lattice constants are fixed,
the in-plane stress will change with electric field.
We use the thermodynamic relation:
\begin{equation}
  \eta_p = S_{pq}\sigma_q + d_{pm}E_m
  \label{eqn:etasigmaE}
\end{equation}
where $S$ is the fixed electric field compliance tensor and $\eta$ is the
strain in Voigt notation\cite{fundamentalsPiezo}.
With zero in-plane strain ($\eta_1=\eta_2=0$), $\sigma_3=0$, and using the tetragonal symmetry of the
systems examined in this work we can obtain from equation (\ref{eqn:etasigmaE}) 
\begin{equation}
   \sigma_1 = \sigma_2 =  -\frac{d_{13}}{S_{11} + S_{12}} E_3
  \label{eqn:sigmaE}
\end{equation}
Next we utilize the thermodynamic relation:
\begin{equation}
  D_m = \epsilon_{mn}E_n + d_{pm}\sigma_{p}
  \label{eqn:DEsigma}
\end{equation}
where  $\epsilon$ is the zero-stress
dielectric tensor and $d_{pm} = (\frac{d D_p}{d \sigma_m})_{E=0}$ is the
zero-stress piezoelectric tensor,
and we differentiate $D_3$ with respect to $E_3$ obtaining:
\begin{equation}
  \frac{dD_3}{dE_3} = \epsilon_{33} + \sum_i(d_{i3}\frac{d\sigma_i}{dE_3})
  \label{eqn:dDdE}
\end{equation}
The $\frac{d\sigma_i}{dE_3}$ are easily obtained from (\ref{eqn:sigmaE}) and inserted into the above expression
 to obtain the desired epitaxially-constrained dielectric permittivity:
\begin{equation}
  (\frac{dD_3}{dE_3})_{\sigma_3=0} = \epsilon_{33}  - \frac{2 d_{13}^2}{S_{11} + S_{21}}
  \label{eqn:epi_eps}
\end{equation}

Now we turn to the epitaxially-constrained piezoelectric response
$(\frac{dD_3}{d\sigma_3})_{E=0}$.
To express this in terms of the zero-stress quantities provided by
ABINIT, we proceed in close analogy to the discussion for $\epsilon_{33}$ above.
Note that with in-plane strain fixed, in-plane stress will change as $\sigma_3$ is varied.
Again using thermodynamic relation (\ref{eqn:etasigmaE}), still with
$\eta_1=\eta_2=0$ and tetragonal symmetry, but now with $E_i=0$ for all $i$, we
can obtain
\begin{equation}
\sigma_1  = \sigma_2 = -\frac{S_{13}}{S_{11} + S_{12}} \sigma_3
  \label{eqn:sigmasigma}
\end{equation}
Making use of the thermodynamic relation (\ref{eqn:DEsigma}) we differentiate
$D_3$ with respect to $\sigma_3$ obtaining
\begin{equation}
  \frac{dD_3}{d\sigma_3} = \sum_i d_{i3}\frac{d\sigma_{i}}{d\sigma_{3}}
\end{equation}
The desired $d\sigma_i / d\sigma_3$ are easily obtained from (\ref{eqn:sigmasigma})
yielding the
epitaxially-constrained piezoelectric response:
\begin{equation}
  \frac{dD_3}{d\sigma_3} = d_{33} - \frac{2 d_{13}S_{13}}{S_{11} + S_{12}}
  \label{eqn:epi_d}
\end{equation}

The quantities $d_{33}$, $d_{13}$, $S_{13}$, $S_{11}$, and
$S_{12}$ that appear in the right hand side of equations (\ref{eqn:epi_eps})
and (\ref{eqn:epi_d}) can be obtained in a
straightforward manner using the DFPT
implementation in ABINIT along with the ANADDB post-processing tool.
This is also true for $\epsilon_{33}$, so long as
the system contains no unstable phonon modes at the $\Gamma$ point.
However, in some ${\mathrm{PbTiO_3} / \mathrm{SrTiO_3}}$ $1x1xN$ $P4mm$
superlattice structures we find unstable $E_u$ polar modes meaning the full
dielectric permittivity matrices can not be obtained directly using ANADDB.
However, by symmetry the oscillator strengths of these modes are such that they do not
contribute to $\epsilon_{33}$.
In this case we obtain $\epsilon_{33}$ using quantities that are output from
ABINIT and ANADDB,
first computing the zero strain $\epsilon_{33}$ from equation 53 from
\cite{PhysRevB.55.10337} and then obtaining $\epsilon_{33} =
\epsilon^{(\eta=0)}_{33} + \sum_p e_{p3}d_{p3}$, where $e_{ij} = d D_{j} / d
\eta_{i}$ and $\eta$ is the strain in Voigt notation. 

\section{Comparison of linear response and finite field results for bulk
constituents}
\begin{table}
 \begin{minipage}{0.3\linewidth}
 \begin{tabular}{|l | r r r|}
 \hline
 \multicolumn{4}{| c| }{$\mathrm{SrTiO}_3$} \\
 \hline
 & $d_{33}$ &$\epsilon_{33}$ & $g_{33}$ \\
 \hline
 FF & 69.0 & 510 & 0.01528 \\
 LR & 71.5 & 528 & 0.01530 \\
 diff & 2.5 & 18 & 0.00002 \\
 \% diff & 3.6 & 3.4 & 0.2 \\
 \hline
 \end{tabular}
 \end{minipage}
 \begin{minipage}{0.3\linewidth}
 \begin{tabular}{|l | r r r|}
 \hline
 \multicolumn{4}{| c | }{$\mathrm{BaTiO}_3$} \\
 \hline
 & $d_{33}$ & $\epsilon_{33}$ &
$g_{33}$ \\
 \hline
 FF & 31.6 & 38.1 & 0.0936 \\
 LR & 30.9 & 37.2 & 0.0937 \\
 diff & 0.7 & 0.9 & -0.0001 \\
 \% diff & 2.3 & 2.4 & -0.1 \\
 \hline
 \end{tabular}
 \end{minipage}
 \begin{minipage}{0.3\linewidth}
 \begin{tabular}{|l | r r r|}
 \hline
 \multicolumn{4}{| c | }{$\mathrm{PbTiO}_3$} \\
 \hline
 & $d_{33}$ & $\epsilon_{33}$ &
$g_{33}$ \\
 \hline
 FF & 56.8 & 62.6 & 0.1026 \\
 LR & 58.0 & 63.7 & 0.1029 \\
 diff& 1.2 & 1.1 & 0.0003 \\
 \% diff& 2.1 & 1.8 & 0.3 \\
 \hline
 \end{tabular}
 \end{minipage}
 \caption{Comparison of finite field (FF) and linear response (LR) results
   for the dielectric permittivity ($\epsilon_{33}$) and piezoelectric responses
   ($d_{33} = \frac{1}{c}\frac{dc}{dE}$ and $g_{33} = \frac{1}{c}\frac{dc}{dD}$) of each
   epitaxially constrained material.}
 \label{table:dfpt_field}
\end{table}
In Fig. 2 of the main text we noted that the end point linear response calculations do not coincide
perfectly with the model curve. In the DFPT calculations the response of the
material to an electric field is computed using derivatives of the wavefunction
with respect to the wavevector ($\vec{k}$) in the Brillouin zone (BZ), which are
found by solving a Sternheimer equation at each k point.
In the finite field calculations used to parameterize the model, dependence on
the wavefunction on wavevector ($\vec{k}$)
across the BZ are instead incorporated through the
polarization  ${P}$ term in the energy functional. These two methods
converge differently with respect to k-point sampling and plane wave basis \cite{Wang_2007, Wu_2005} 
resulting in small differences between the model and DFPT results even for bulk
compounds; details are given in table \ref{table:dfpt_field}.

\bibliography{dielectric_slab}{}
\end{document}